# Human Aspects and Perception of Privacy in Relation to Personalization

Sanchit Alekh

Informatik 5, Databases and Information Systems
Chair Prof. Dr. Stefan Decker
RWTH Aachen University
Aachen, Germany

**Abstract.** The concept of privacy is inherently intertwined with human attitudes and behaviours, as most computer systems are primarily designed for human-use. Especially in the case of Recommender Systems, which feed on information provided by individuals, their efficacy critically depends on whether or not information is externalized, and if it is, how much of this information contributes positively to their performance and accuracy. In this paper, we discuss the impact of several factors on users' information disclosure behaviours and privacy-related attitudes, and how users of recommender systems can be nudged into making better privacy decisions for themselves. Apart from that, we also address the problem of privacy adaptation, i.e. effectively tailoring Recommender Systems by gaining a deeper understanding of people's cognitive decision-making process.

## 1 Introduction

With the success of internet-based services such as online marketplaces, social networking, digital media streaming et cetera, the use of recommender systems is becoming increasingly widespread. Almost all the technology giants today, are using recommender systems in one form or the other [1, 2]. On the other hand, at the receiving end, the number of users of these systems has increased several-folds in the previous decades [3], according to the International Telecommunication Union. While there has been a considerable volume of high-quality research in the area of automatic recommendation techniques [4, 5], there is a common consensus among privacy scientists that recommender systems algorithms are only a small contributing factor in the overall user-experience [6]. This is primarily because recommender systems heavily rely on the end-users to externalize their personal data, and the efficacy of these algorithms depends on the accuracy and relevance of this information. The concept of privacy, as well as the act of data disclosure are human behaviours, and developers of recommender systems must conduct in-depth experiments about the users' privacy-related attitudes and their information disclosure behaviour to augment user-satisfaction [7].

### 1.1 Motivation

Massive amounts of user data is generated in the scope of recommender systems. The ubiquity of recommender systems, their utility and user base makes them a threat to user privacy, if the data is inappropriately accessed or used. Therefore, it is imperative for system architects and designers to find a balance, so that they can not only generate accurate recommendations, but also guarantee privacy of their users [7]. In this regard, several algorithmic as well as cryptographic approaches have been suggested, including pseudonymity-based personalization [8, 9], obfuscation [10, 11], differential privacy [12] and additive homomorphic encryption using Paillier public-key cryptosystem [13].

These approaches are, however, purely technical in nature, and do not take into account the end-users, for whom privacy is intended to be championed. Indeed, Friedman et al. [7] argue that in most cases, the users' cognitive resources are insufficient to effectively take control over their privacy. Several companies and their recommender systems implementations make use of this dichotomy in privacy attitudes and behaviours of users, in order to extract more information from the user to make their systems more effective. This dichotomy is known as *Privacy Paradox* in privacy literature, and was first pointed out by Norberg et al. [14]. Therefore, it is important to better understand the privacy needs of individual users, provide them with transparent options, and help them make the right privacy decisions for themselves.

### 1.2 Outline of the Paper

In the next section, we talk about privacy attitudes and behaviour. In particular, we focus on the dichotomy between the two, also known as *Privacy Paradox*. In *Section 3*, we talk about the internal and external factors that influence information disclosure and decision making in an online scenario. In *Section 4*, we bring together inferences from studies in the area of privacy and information systems research and psychology to suggest steps and techniques that developers and companies building recommender systems should follow while designing such systems. These best practices would ensure that the user's privacy concerns are placed at the center of the system, and their privacy is not undermined.

## 2 Privacy Attitudes and Behaviour

### 2.1 Privacy Paradox

Norberg et al. [14], in their studies, essentially investigate the factors of *risk* and *trust* with respect to actual human privacy behaviours, and the relationships between these two factors. They wanted to ascertain whether risk considerations also induce a user's actual information disclosure behaviour. These studies were motivated by the fact that the onset of the information age has enabled collection, sharing and processing of huge amounts of data, and has given marketers access to a treasure of personal user data and preferences. However, this has, at the same time, serious ramifications for user privacy. The authors argue and hypothesize that *risk* influences a user's intention to disclose but *risk factors* are not strong enough to influence actual disclosure behaviour. In other words, when a user is directly asked about his *intention* to disclose, risk factors are at play, however, in an actual *disclosure* scenario, *trust* as an environmental cue is expected to significantly impact the user's decisions.

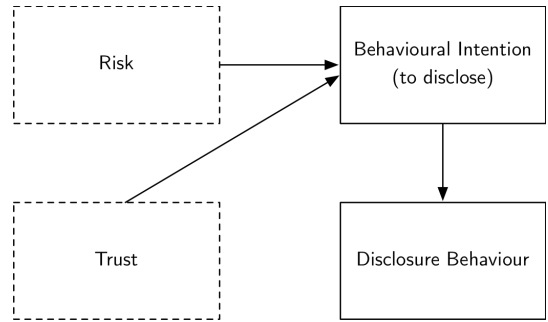

**Fig. 1.** The relationship between intentions and behaviour: Traditional Approach

Previous research [15,16] has focused on the interfusion of risk and trust in context of a user's behavioural intention. The traditional model says that both risk and trust influence behavioural intentions, which then influences actual behaviour. This is shown in *Figure 1*. However, Norberg et al. argue that behavioural intention is not predictive of actual disclosure behaviour, because *risk* and *trust* operate independently. Detailed user studies to confirm *Privacy Paradox* are well-described in the paper by Norberg et al. [14]

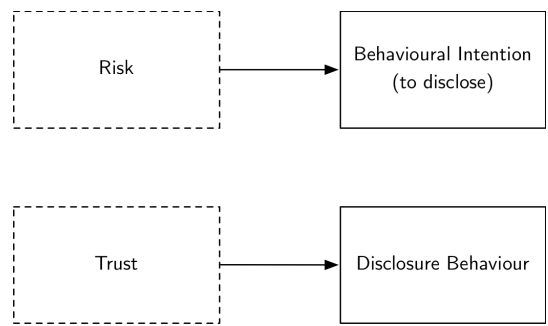

**Fig. 2.** The relationship between intentions and behaviour: Privacy Paradox

## 2.2   Risk and Privacy Attitudes

It has been very well established that behavioral intentions are a culmination of an individual's attitude, subjective norms and perceived control of an outcome [17]. Moreover, O'Keefe and Daniel [18] show that the correlation between intention and behaviour varies between 0.41 and 0.53, and that the measure is more accurate when behaviour is voluntary. Apart from that, the following factors have also been shown to influence the strength of correlation: a) the degree of correspondence between the measure of intention and the measure of behavior [19], b) the temporal stability of intention [19, 20] and c) the degree to which the behaviour was planned [20, 21].

One of the first research efforts in the area of *general privacy concern* was undertaken by Westin and Harris and Associates [22], who classified people into three categories based on their privacy attitudes: *Privacy Fundamentalists*, *Pragmatists* and *Unconcerned*. However, in the context of user privacy, it is indeed evident that one's stated intentions might not be reflective of behaviour, because of factors like *heuristic processing* and *routinization of behaviour*, that influence intention and behaviour very differently [14]. Moreover, it has also been shown that *social desirability bias* could contaminate one's response to the question of *behavioural intention* to an extent that it would no longer be indicative of actual behaviour.

Moreover, the most important consideration for recommender systems designers is that privacy concerns are extremely system/context-specific, rather than a generic factor [23]. Therefore, a user's privacy attitude can be significantly different based on system-specific factors such as a) *perceived* privacy threats [9, 24], b) *perceived* protection [9] and c) trust in the company [24], among others. Therefore, system engineers, architects and developers should make sure that they focus not only on the protecting the user privacy via technical means, but try to make the user feel that they are at the center of the system, i.e user privacy is of paramount importance at every stage of the system. This philosophy is often referred to as *Privacy by Design* [25].

At this point, it might be worthwhile to excurse and talk about *Privacy by Design*, introduce the idea and concept behind it, and what it means and represents.

**Privacy by Design** Cavoukian [25] lays out 7 foundational principles of Privacy by Design(PbD). PbD is a set of guidelines for software system architects to design privacy-centred systems. According to her,

> The universal principles of the Fair Information Practices (FIPs) are affirmed by those of Privacy by Design, but go beyond them to seek the highest global standard possible. Extending beyond FIPs, PbD represents a significant "raising" of the bar in the area of privacy protection.

The 7 foundational principles of PbD are: a) Proactive not Reactive; Preventative not Remedial, b) Privacy as the Default, c) Privacy Embedded into Design,

d) Full Functionality - Positive-Sum, not Zero-Sum, e) End-to-End Security - Lifecycle Protection, f) Visibility and Transparency, and g) Respect for User Privacy

### 2.3 Trust and Privacy Behaviour

Several psychologists have noted that while making privacy decisions, people often make a trade-off between risks and benefits [26]. Culnan and Bies [27] call this as *Privacy Calculus*. Privacy Calculus is also influenced by several other factors, but the most influential of them is *trust*. Trust can be defined in several ways, one of them is 'willingness to rely on an exchange partner' [28]. Trust can be tangibly measured in terms of a company's reputation, brand name and status. This means that in spite of the actual trade off between risks and benefits, a user is expected to disclose more information if he places a considerable amount of trust in a company. Cranor et al. [29] have indeed shown that there is a positive correlation between the amount of information disclosed, and the amount of trust reposed in a particular brand or company.

To motivate this, let us take an example with two hypothetical companies *Doogle* and *BuckBuckGo*, both of which are into the business of recommending restaurants. The user has been using products offered by *Doogle* for a long time, and holds the company in high regard. On the other hand, *BuckBuckGo* is a relatively new entrant, and doesn't offer any other services apart from restaurant recommendation. Both the companies ask for the user's home address as a part of the service sign-up process. The user is quite privacy-minded, i.e. a *Privacy Fundamentalist*, according to the classification criteria of Westin and Harris and Associates [22], i.e. he/she believes that one should not disclose their home address in an online scenario. The research of Cranor et al. [29], however, posits that nevertheless, there is a very high probability that the user in fact, discloses his/her home address in order to use the service offered by *Doogle*. However, the same user might be inclined to think think that their home address is not directly related to the service being offered by *BuckBuckGo*, and therefore would avoid using that service. This is also confirmed and reiterated by Norberg et al. [14] in their research on *Privacy Paradox*. This example clearly demonstrates that a person, whilst having the same privacy attitudes, shows different disclosure behaviour.

To conclude, privacy paradox exists and is very much apparent, e.g. in the online marketplace. This means that users are often found making a trade-off between risks and benefits of disclosure, also known as *Privacy Calculus*. This decision is also dependent on the context and relevance [7] with respect to the purpose of the system. Another factor, which consequently operates, is the amount of trust that the users repose in that brand or company.

## 3  How do humans perceive Privacy in the context of Personalization?

In the previous section, we have described the intimate ways in which risk and trust act to determine privacy attitudes and behaviour. But the question arises, what factors do privacy decisions really depend on? This question can only be answered by a close study into human psychology. The research performed by Acquisti et al. [30] systematically lays down three interconnected factors which influence decision-making in privacy scenarios: a) People's uncertainty about the consequences of privacy-related behaviors and their own preferences over those consequences b) The context-dependence of people's concern, or lack thereof, about privacy, and c) The degree to which privacy concerns are malleable, i.e. manipulable by commercial and governmental interests. We explain in detail, each of the factors in the subsections below.

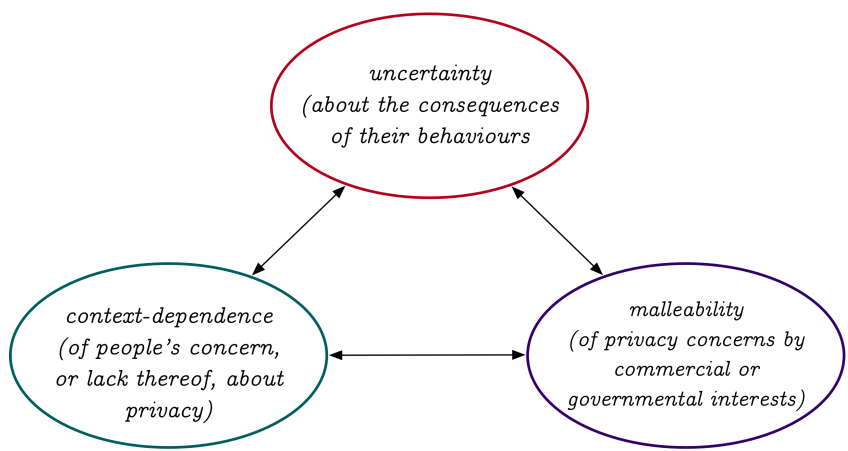

**Fig. 3.** Factors which affect Information Disclosure. The double headed arrows signify that these factors are connected to each other, and influence each other

### 3.1  Uncertainty

According to Acquisti et al. [30], the foremost and most obvious form of privacy uncertainty stems from incomplete or asymmetric information. In most instances, users of recommender systems do not have a clear knowledge of what data has been collected from them, and how it is being used, or would be used in the future. The frequently changing End User License Agreements (EULA), along with the fact that they are worded in such a way that makes a normal user averse to reading them, also contributes to this problem. Therefore, people are uncertain about how much information they should disclose. The authors also say that whereas on one hand, some privacy harms are tangible, such as financial costs due to identity theft, several others, such as a voyeur getting access to someone's private images, are intangible. The second reason for uncertainty relates to *Privacy Paradox*. People are extremely sure of their privacy preferences, although they are unaware of the consequences of their behaviours. In this regard, Singer et al. [31] conducted a study in which they asked sensitive and potentially incriminating questions either point-blank, or accompanied with assurances of confidentiality. Contrary to expectation, these assurances did not lead to broader divulgence, because they elevated the users' privacy concerns. People's privacy attitudes are made complicated by the existence of a powerful countervailing motivation: the desire to be public, share, and disclose. Acquisti et al. [30]argue that humans are social animals, with an inherent desire to share personal information. Information sharing is a central feature of human connection. This is also confirmed by social penetration theory [32], which says that increasing levels of disclosure are natural, and a desirable evolution of interpersonal relationships from superficial to intimate.

### 3.2  Context-Dependence

Context plays a major role in determining how much information is disclosed in what kind of a situation. Acquisti et al. [30] find that when people are unsure about their preferences in a particular situation, they take cues from the environment to aid them to make a decision. To speak in terms of the classification by Westin [22], humans are *Privacy Fundamentalists, Pragmatists* or *Unconcerned* based on time and place. There are several individual factors, which together embody context-dependence.

One of them is the amount of closeness with the person with whom information is being shared. E.g. it might be much easier to share sensitive information with a friend rather than a stranger on the flight [33]. In the context of personalization and recommender systems, this translates to brand confidence and familiarity with the services offered by the company. Moreover, stringent goverment regulations are a favourable factor, enabling people to share more information, as they reduce concern and increase trust [34]. One startling and counter-intuitive factor is user-interface design. John et al. [35] found in an online study, that people were much more likely to disburse personal and incriminating information on an unprofessionally and casually designed website with a banner *'How Bad R U?'*, compared to a website with a much more formal interface. Yet another contextual factor is the physical environment encompassing the person. It has been shown by Chaikin et al. [36] that all other factors being equal, intimacy of self-disclosure is larger in warm and comfortable surroundings, with soft lighting rather than in cold rooms with bare cement and fluorescent lighting. It has also been shown that people are influenced to externalise more personal information when they get to know that other people have also given out the same information. *Turbulence of Privacy Boundaries*, i.e. when trust boundaries are broken, e.g. a post on social media reaching an unintended audience, people are much more likely to exercise restraint and limit the visibility of their future posts to a selected audience.

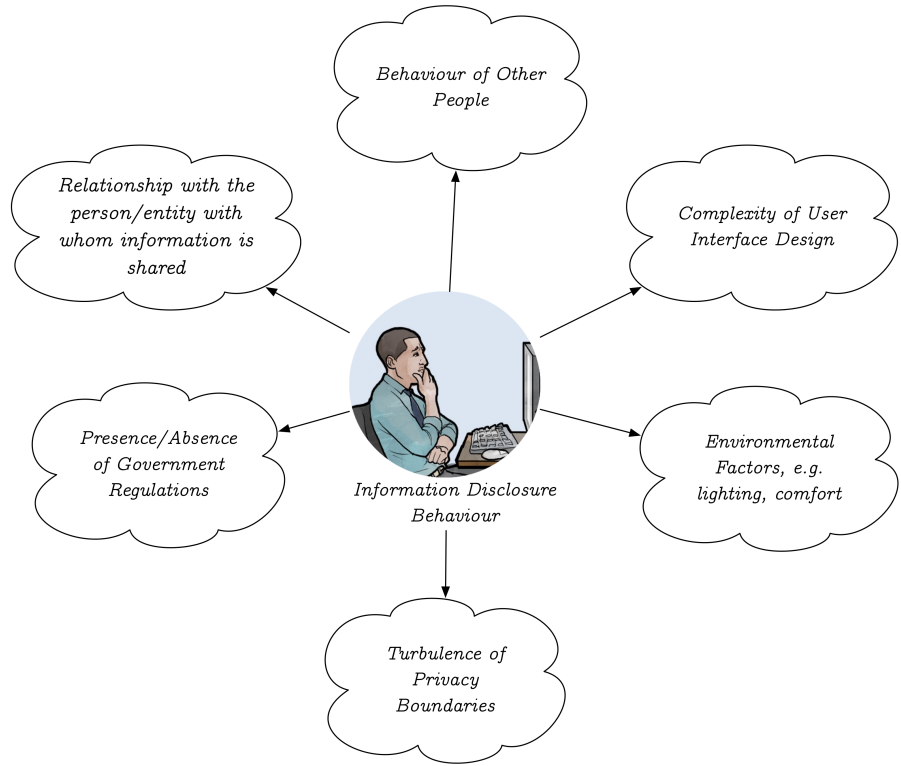

**Fig. 4.** Context Dependence of Privacy Decisions

All these criteria directly apply to systems which aim for personalization, because they are all dependent on a user disclosing personal, potentially-incriminating information such as shopping history, hobbies, age, birthday and so on.

### 3.3 Malleability

The tremendous amount of available data has given rise to entities whose economic or business interests are to gain personal information. Companies that build recommender systems are counted among such entities, apart from several other types of businesses, e.g. those involved in behavioural advertising or online social networks. These entities study behavioural and psychological processes and incorporate subtleties which affect the user's disclosure behaviour. This leads to *malleability* of the user's privacy preferences. *Defaults* are a common and widely used example. Defaults are considered by users as implicit recommendations, and it is often much more convenient for people to let the default option be their choice. This has been used in websites collecting *organ donor* or *retirement savings* information, and also in privacy settings for *social network profile visibility* [30]. Another such manoeuvre is *Malicious Interface Design*, by which entities develop their user interface features in a way that frustrates or irritates a user into providing more information. A research by Hoofnagle et al. [37] also showed that merely showing or posting a *Privacy Policy*, which users might not even read, led to users having a misplaced feeling of being protected. Another strategy that service providers knowingly use, is to not alert the user, or send a warning, when a potentially harmful privacy setting is used.

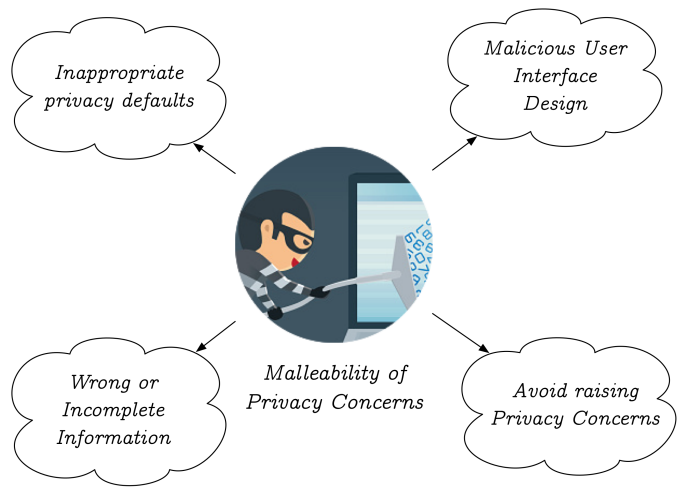

**Fig. 5.** Malleability: Factors which seek to undermine the user's privacy decisions

## 4 Towards a Human-Centric Privacy Adaptation

We have looked at various human factors and tendencies in regards to information disclosure, especially in the context of personalized systems. We have also looked at how humans perceive privacy in such systems, and how internal and external factors influence decision making. Recommender Systems and other such systems which aim for personalization have definitely proved to be a boon for internet users. However, this has come at a cost, i.e., users have lesser control over personal information that the companies know about them.

Companies and developers of recommender systems often take advantage of inherent *human weaknesses* to cajole system users to externalize more information. At the same time, there are several start-up companies as well as entities trying to foray into personalized systems, looking for best practices for developing a system that is extremely useful, but at the same time, designed with user privacy in mind, and that places the user at the center of the system. Such a set of guidelines is not readily available to guide the developers. *Privacy by Design* [25] is a step in the right direction, however it is often too generic, vague and idealistic to provide realistic guidelines of any kind.

In this section, we try to introduce concepts and best practices which enable the development of *Human Centric Personalization Systems*. This makes sure, that the users have an accurate knowledge of what the systems aim to provide, and what they, in turn, expect from them. At the same time, most users of these systems are not specialists in computer science, psychology or law. Therefore, the users' cognitive resources are in most cases, insufficient to effectively take control over their privacy and make safe and informed decisions [7]. Therefore, the onus falls on the entities developing these systems to *guide* the users make the best privacy decisions for themselves. This is a win-win situation for both parties, because it has been shown that building trust reduces concerns about privacy, and the user discloses more information. At the same, it portrays a positive image of the company in the minds of the users.

### 4.1 Striking the correct balance of Transparency and Control

Both transparency and control are fundamental to gaining the user's trust and helping them engage in a privacy calculus. People can make informed decisions only if adequate information is provided to them. It has often been suggested that recommender systems should give users advanced controls over their privacy. However, while users often claim to want complete control of their privacy, they eschew the trouble of exercising this control [38]. A fully transparent system suggests that all the fine-grained, ground details and information regarding their privacy should be available to the user. This is not a bad thing to do, but it has been shown to hinder system usage rather than encouraging it [39]. Bakos et al. [40] found that only a miniscule 0.2% people read End User License Agreements (EULAs), and summarizing them with trust seals is often counter-productive [39]. This is not surprising, as most of these documents are lengthy and filled to the brim with legal and technical jargon. Although it shows positive intent on part of the company, it often does not fulfill its objective. Keeping all the above factors in mind, we would like to suggest the following best practices to achieve the right amount of transparency and control.

– Displaying the complete EULA is not bad and indeed, in most cases, it is also a legal requirement. However, there should be a real effort to explain to the user, the contents of the EULA which are *most relevant for their privacy*. This should be done in way that the user can *easily understand and absorb*, e.g. with a video or animation. The user should preferably get the idea that a human is talking to them, and if possible, with a two-way conversation and not just a monologue. Recent advances in speech recognition and virtual assistants have made this possible [41]. This feels much more intuitive to the user, and builds trust easily, while also making them aware of the fundamental privacy aspects of the system.

- A lot of times, the user does not have the time to go through all the privacy aspects of the system during the signup process. Therefore, this should be made available at all times for *future reference*, so that they have the option of having a look at it at a later stage.

- Future modifications of the EULAs and privacy policies of the companies should avoid relaxing currently existing ones. If this is unavoidable however, the user should at least be well-informed about the important changes, and should be provided with an alternative, if they would like to opt-out of providing personal information due to the ramifications of these changes.

- All privacy controls relevant to the system should preferably be condensed into one page, e.g. in a *Dashboard* format, to make sure that exercising control over privacy is convenient and not distressing to the user. Moreover, these settings should be easily accessible, and not obscure or confounding.

- To especially cater to the needs of technically oriented or advanced users, it should also be possible to know about the fine-grained details of exactly what information is collected from the users, and how it is used to make the recommendations better. These descriptions could be technical in nature, and need not necessarily be shown to the users during the signup process, but available for reference. This would make sure that it doesn't flare up anxiety in the minds of the user, but at the same time, the system remains fully transparent.

### 4.2  Privacy Nudges

An important feature that developers should incorporate into their personalization systems is called as *Privacy Nudges*. Nudges are meant to drive users towards making the *right* privacy decisions for themselves. Recent research [42] has shown that nudges are quite often a very effective technique that helps reduce the regret associated with decision-making, and that helps users make the correct choice without limiting their ability to choose freely. [7]. Privacy Nudges can be divided into two broad categories: *Justifications* and *Defaults*.

**Justifications** *Justifications* are an effective way to help a user rationalize their decisions. It also makes them better understand the significance of why the information is required by the system, and how it makes the user experience better. There can be different ways of providing justifications, some of them are: a) Providing a reason for requesting the information [43]  b) Highlighting the benefits of disclosure [44], and c) Appealing to the social norm [45, 46] . Again, these justifications need to be made in a manner that is not too intrusive, and that blends well with the design of the system, rather than appearing to be an erroneous parameter. At the same time, they should also be easily understandable, even by the non-expert user. The actual results produced by using justications seems to be inconclusive. Acquisti et al. [46] found that users were 27% more likely to disclose the same information, when they learnt that other people had also agreed to disclose that information. On the other hand, Knijnenburg et al. tested justifications in a demographics and context-based mobile app recommender [47], and found that it actually decreased disclosure. However, most studies agree that users find justifications to be helpful in making privacy-related decisions, and that only a particular subset of users are amenable to justifications, i.e. non-expert users [45].

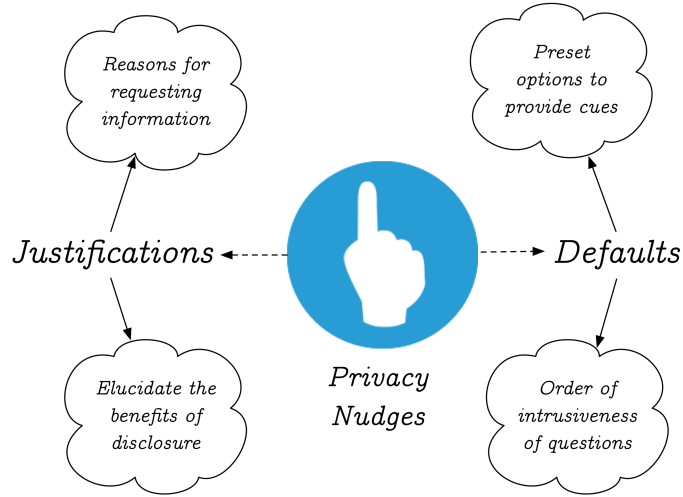

**Fig. 6.** Privacy Nudges: Guiding the users towards making the right privacy decisions for themselves

**Defaults** *Defaults* are another means of nudging a user. They aim to release the user's burden of having to make a decision by himself. Defaults are essentially *implicit recommendations*, and they play a significant role in influencing privacy decisions [48]. Carefully setting defaults can be a stimulant, and can increase user participation. However, the defaults should make sure that user privacy is not undermined. Another form of defaults is the order of intrusiveness in which questions are posed. Acquisti et al. [46] have shown that posing questions in a random order of intrusiveness is more effective than posing them in an increasing order of sensitivity.

### 4.3   Tailor-Made Privacy Decision Support: Privacy Adaptation

In *Section 4.2*, while introducing *Privacy Nudges*, we have asserted that processes such as *justifications* and *defaults* are meant to nudge the user towards making the *right* decision. With this description, arises an extremely plausible interrogation: *What is meant by the "right" decision?* We have already seen, especially in *Section*

*3.2* that human decisions are extremely context-dependent, and this applies to information disclosure decisions as well. Internal and external factors play a pivotal role in determining what information is externalized. Keeping this into account, it can be argued that no single privacy default can be ideally suited for all users. What is ideal for one user, might not be ideal for another [49]. The same holds for justifications as well, i.e., what is justifiable for one user, might be completely irrelevant for another user, depending on the item to be disclosed and different contextual factors [7, 45]. These arguments pave way for the application domain of *Tailor-Made Privacy Decision Support* for recommender systems. This is also called as *Privacy Adaptation*, i.e. adapting privacy nudges for an individual user by modeling their behaviour in different disclosure scenarios.

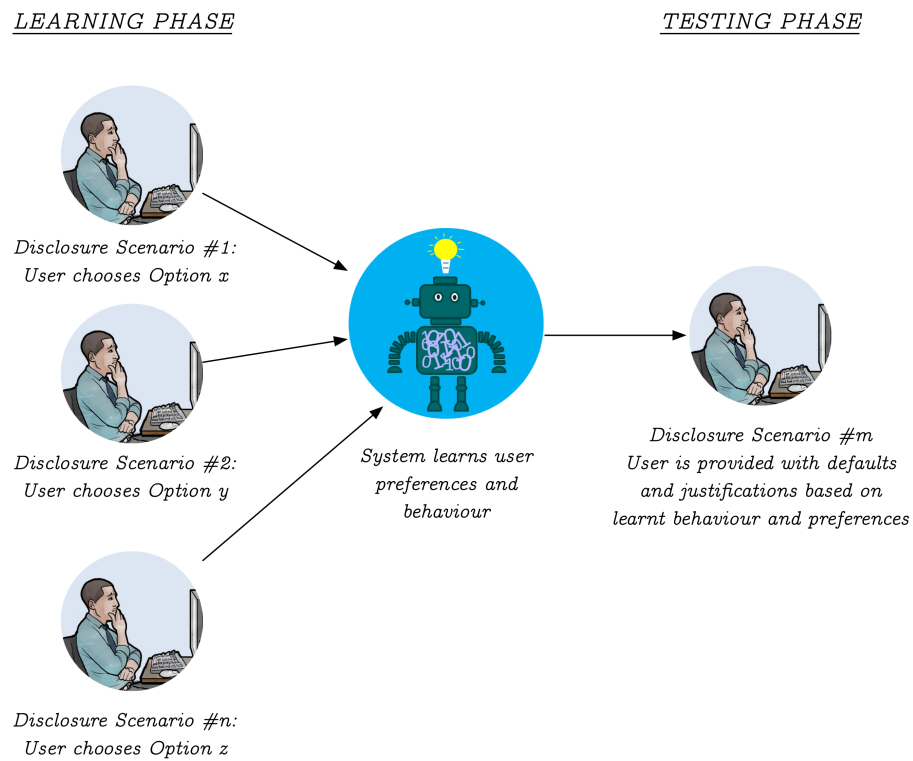

**Fig. 7.** A brief overview of the steps involved in Privacy Adaptation

Privacy Adaptation provides a tenable middle-ground between giving full privacy control to the user, and giving them no control at all. Gaining a deeper understanding of a user's cognitive decision-making behaviour not only helps the system to suggest better privacy defaults, but enables the user to repose more faith and trust in it [24]. The work of Knijnenberg et al. has focused on trying to measure these information disclosure decisions and materialize them into behavioural models of information disclosure decisions. They take into account the *Privacy Calculus* that the user is confronted with, i.e. the balance of risks and benefits for the user in a particular disclosure scenario [50]. Apart from this, the other factors are the recipient of the information, the item to be disclosed, and the kind of user (i.e. privacy attitude). They also try to group users based on their privacy preferences in several domains, and identify distinct subgroups [49]. These subgroups, when combined with other information such as demographics and other behaviours, can be effectively used to tailor privacy decisions for the user. Similar efforts have been made by Ravichandran et al. [51], who have tried to use k-means clustering to cluster users in the context of their location sharing preferences, and devise default policies based on this. They have also shown that a small number of default policies are able to represent a large number of users, which significantly reduces the complexity of tailoring privacy decisions. Sadeh at al. [52] also tried to learn user privacy preferences in a location-sharing system by applying the k-nearest neighbour algorithm and random forests. Schaub et al. [53] have proposed a framework for dynamic privacy adaptation system in the context of ubiquitous computing.

In effect, Privacy Adaptation can be succinctly described as *a recommender system to enable user privacy in a recommender system*. Although our previous examples describe the application of privacy adaptation for tailoring privacy defaults, they can also be used for customizing justifications. Although we have shown in *Section 4.2* that justifications do not necessarily lead to an increase in information disclosure, customizing justifications based on user preferences counterbalances their negative effect [54].

Privacy Adaptation provides a unique way to find a balance between too much and too little control for the user. It gives us a way to find the *"right"* privacy decision for the user by using their own preferences as a yardstick. Context-aware approaches also provide an effective solution to the problem when the user is overwhelmed by too much information, which further heightens their privacy concerns.

## 5   Conclusion and Discussion

In this paper, we have tried to argue that technical solutions are not enough to address privacy concerns and challenges in the domain of personalized systems. These systems are developed for human consumption, and therefore, psychological and behavioural issues are extremely important to be taken into account. We have shown how a user's privacy attitudes are not a true reflection of their actual disclosure behaviour. This is mainly because they are driven by totally independent factors, that of *risk* and *trust* respectively, which operate in different ways. For this reason, we consider the user's behaviour and preferences in actual online disclosure scenarios for building adaptive privacy recommending systems. In *Section 3*, we have presented and described the various factors that influence the user's decisions. A close study of these factors is essential not only for developers of recommender systems, but also sometimes for users, to be better aware of how their inherently "human" characteristics and traits can be leveraged by recommender systems to make them disclose more information than they actually intended to.

We have discussed at length, not only about the effects of uncertainty on information disclosure, but also of seemingly innocuous context-factors such as lighting and temperature around the user. In addition, we have shown how companies and other entities can use tricks such as malicious user interfaces to undermine user concerns about privacy. We have presented several techniques and best practices that system developers and start-up companies foraying into recommender systems and allied areas should consider. This is especially essential for gaining a user's trust and reducing the possibility of any untoward privacy related mishaps. The concept of Privacy Nudges has been introduced, by which we aim to help the user make correct privacy decisions for themselves.

Finally, we have shown that the meaning of *"right"* privacy decisions can vary, not only from user-to-user, but also from context-to-context. What kind of information is to be disclosed, who is the receiver of this information, what kind of service is being provided to the user in lieu of this information etc. play a decisive role in determining whether disclosing the information is the *"right"* decision for a user. For this purpose, it is highly recommended to include some form of *Privacy Adaptation* in modern recommender systems. This provides a method to enable user-centric privacy defaults and justifications, which are tailored for the individual user, using his own preferences and behaviours in real information disclosure scenarios. We would like to conclude by asserting that if these recommended proposals are considered by developers and entities building personalized systems, it would take us one-step closer to realizing truly user-centric systems, and would extinguish privacy concerns to a great extent.

All links were last followed on July 13, 2017.